\documentclass{svproc}
\usepackage{graphicx}
\usepackage{mathtools}
\usepackage{amsmath,amsfonts,amssymb}
\usepackage{tikz}
\usetikzlibrary {arrows.meta,bending,positioning}
\usepackage{tikz-dimline}
\usepackage{float}
\usepackage{subcaption}
\usepackage{bm}
\usepackage{multirow}
\usepackage{siunitx}
\usepackage{comment}
\usepackage{listings}
\usepackage{hyperref}
\hypersetup{pdftex=true, colorlinks=true, breaklinks=true, linkcolor=blue, menucolor=blue, citecolor=blue, urlcolor=blue}
\usepackage{xcolor}
\usepackage{algorithm}
\usepackage{algorithmic}
\definecolor{beaublue}{rgb}{0.94, 0.97, 1.0}
\usepackage{url}

\begin{document}
\mainmatter              

\title{GO-GAN: Geometry Optimization Generative Adversarial Network for Achieving Optimized Structures with Targeted Physical Properties}

\author{ A. Padmaprabhan\inst{1}, Shriram Hari\inst{1}, Nived Philip Thomas\inst{2}, Khaish Singh Chadha\inst{1}, Sai Sidhardh\inst{1}, Viswanath Chinthapenta\inst{1}, \and Prabhat Kumar\inst{1}}
\titlerunning{GO-GAN: Geometry Optimization Generative Adversarial Network}
\authorrunning{A. Padmaprabhan et. al}
\institute{Department of Mechanical and Aerospace Engineering, Indian Institute of Technology Hyderabad, Telangana 502285, India \\ \and
	Department of Mechanical Engineering, National Institute of Technology Karnataka, Surathkal 575025, India}

\maketitle              

\begin{abstract}
	This paper presents GO-GAN, a novel Generative Adversarial Network (GAN) architecture for geometry optimization (GO), specifically to generate structures based on user-specified input
parameters. The architecture for GO-GAN proposed here combines a  \texttt{Pix2Pix} GAN with a new input mechanism, involving a dynamic batch gradient descent-based training loop that leverages dataset symmetries. The model, implemented here using \texttt{TensorFlow} and \texttt{Keras}, is trained using
input images representing scalar physical properties generated by a custom MatLab code. After training, GO-GAN rapidly generates optimized geometries from input images representing scalar
inputs of the physical properties. Results demonstrate GO-GAN's ability to produce acceptable designs with desirable variations. These variations are followed by the influence of discriminators
during training and are of practical significance in ensuring adherence to specifications while
enabling creative exploration of the design space.
\end{abstract}

\section{Introduction}
Creating intricate geometries in mechanical engineering demands significant computational resources and extensive time. These challenges motivate the exploration of various machine learning and deep learning techniques in the field of mechanical engineering~\cite{shin2023topologyoptimizationmachinelearning}. Abuiedda et al.~\cite{Abueidda_2020} used convolutional neural networks~(CNNs) to optimize nonlinearities in various structures. Rade et al. \cite{Rade_2021} used neural networks to predict compliance and density values, which reduces computation time compared to a conventional topology optimization algorithm. This study explores deep learning, especially generative adversarial networks (GANs, cf.~\cite{goodfellow2014generative}), to optimize and enhance the design process. Its applications include simulation acceleration~\cite{app131910664}, material design, and topology optimization. GAN-based image translation helps optimize mechanical design since it helps with tasks like style transfer and super-resolution while preserving important content across domains.

The paper presents two applications of the proposed GAN model: topology optimization of a cantilever beam for the specified volume fraction and Poisson's ratio and the optimized chair design using the RC49 dataset~\cite{ding2023continuousconditionalgenerativeadversarial}. The model efficiently generates high-quality, performance-focused designs using a conditional GAN (cGAN) methodology~\cite{mirza2014conditional}. The GAN generates an acceptable material layout for the cantilever beam based on the applied load, boundary condition, given volume fraction, and Poisson's ratio. By design specifications, the chair design adjusts to reproduce ergonomic and aesthetic elements from the RC49 dataset.
\begin{figure}[h!]
	\centering
	\begin{subfigure}[t]{\textwidth}
		\centering
		\includegraphics[scale=0.16]{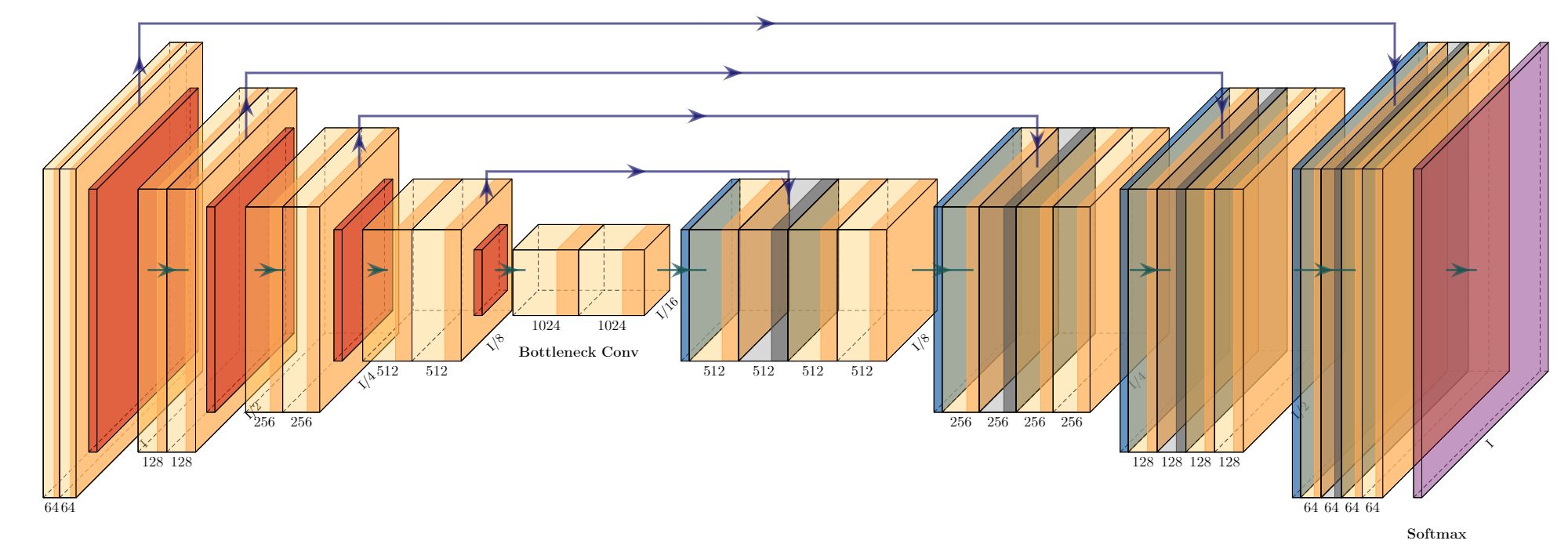}
		\caption{U-Net Architecture}
	\end{subfigure}
	\begin{subfigure}[t]{\textwidth}
		\centering
		\includegraphics[scale=1.3]{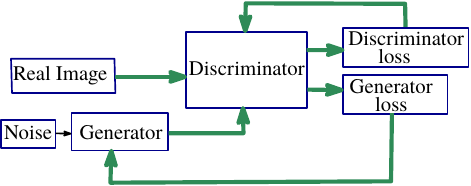}
		\caption{GAN Architecture}
	\end{subfigure}
	\caption{Schematic diagrams for the different architectures used in the proposed neural network model}\label{fig:1}
\end{figure}

U-Net (Fig.~\ref{fig:1}a cf.~\cite{ronneberger2015u}) is a CNN designed for image segmentation, which is effective in topology optimization~\cite{chadha2024pytoacnn}. GANs (Fig.~\ref{fig:1}b cf.~\cite{goodfellow2014generative}) involve a generator and discriminator working adversarially, enabling realistic data generation and driving innovation in design, image generation, and data augmentation. The \texttt{Pix2Pix} GAN, introduced by Isola et al.~\cite{isola2017image}, extends the GAN framework for image-to-image translation using a  cGAN approach with a U-Net-based generator and PatchGAN discriminator. It is considered ideal for tasks like colorization and super-resolution. This framework demonstrates how GANs can automate complex mechanical design tasks by combining binary cross-entropy (BCE) and weighted mean absolute error (MAE) losses with an L1 distance for sharper outputs, advancing structural optimization research and industrial applications while lowering computational costs~\cite{chadha2024pytoacnn}. In addition, the framework includes a customized training approach to meet the requirements for accuracy and structural integrity in the produced designs. The model adjusts to different levels of complexity across design jobs by dynamically modifying hyperparameters like learning rate and batch size during training, producing reliable output quality across applications. The method demonstrates how GANs can produce optimized designs that meet predetermined performance standards and foster creativity by effectively examining various design options.

The remainder of the paper is organized as follows. Sect.~\ref{Sec:2} provides the methodology, detailing the two design tasks--cantilever beam topology optimization and optimized chair design using the RC49 dataset--each defined by specific performance needs. The network architecture, comprising a U-Net generator, PatchGAN discriminator, and combined loss functions, enables high-fidelity design generation. Sect.~\ref{Sec:3} presents results for the model's effectiveness with visual and performance comparisons and discusses the results. Lastly, conclusions are provided in Sect.~\ref{Sec:4}, highlighting its potential, current limitations, and future directions for advancing automated design in engineering.

\begin{figure}[h!]
	\centering
	\begin{subfigure}[t]{0.45\textwidth}
		\centering
		\includegraphics[scale=0.72]{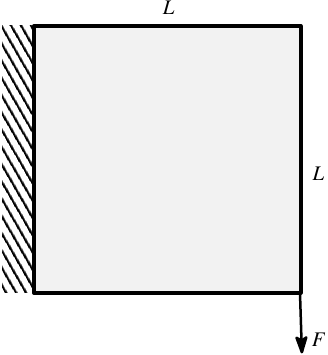}
		\caption{}
	\end{subfigure}
	\begin{subfigure}[t]{0.45\textwidth}
		\centering
		\includegraphics[scale=0.5]{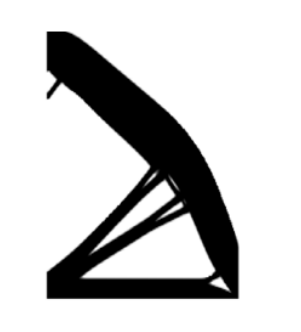}
		\caption{}
	\end{subfigure}
	\caption{Cantilever beam and corresponding optimized design. (a) Cantilever beam design domain. $F$ and $L$ denote the applied load and dimension of the domain. (b) Optimized design obtained using a MATLAB code~\cite{andreassen2011efficient}}\label{fig:your_label}
\end{figure}
\section{Methodology} \label{Sec:2}
This section provides problem descriptions, proposed neural network architecture, input mechanisms, objective loss function, optimization, and algorithm. 
\subsection{Problem Description}
To showcase the GAN's capability in automating various topology and geometric optimization challenges, we evaluate and verify its performance on compliance optimization problems with constant loads~\cite{kumar2023honeytop90} and chair design prediction using the RC49 dataset~\cite{ding2023continuousconditionalgenerativeadversarial}.
\paragraph{Topology Optimization of Cantilever Beam}:
Here, we predict the optimized geometric design of a cantilever Beam (Fig.~\ref{fig:your_label}) by minimizing the strain energy when the load is acting on the end for a given volume fraction and Poisson's ratio~\cite{chadha2024pytoacnn}. The left edge of the beam is fixed, and a load is applied at the right corner of the bottom edge, as shown in Fig.~\ref{fig:your_label}. Optimized beam designs are obtained for training the GAN using the modified SIMP (Solid Isotropic Material with Penalization) approach~\cite{kumar2023honeytop90,andreassen2011efficient}. 
\paragraph{Chair Design Prediction}:
The main challenge in conditional image generation is accurately capturing fine-grained transformations in an object's appearance based on continuous input variables. Small parameter changes, such as object orientation, must result in precise geometric adjustments while maintaining visual coherence. For this purpose, the RC49 dataset~\cite{ding2023continuousconditionalgenerativeadversarial} is used, featuring 3D chair models rendered across yaw angles from $0.1^0$ to $89.9^0$ in fine $0.1^0$ increments. These angles, along with the intended design features of the chair (e.g., armrests, a large backrest, etc.), are used as labels for the corresponding chair configuration. Next, the proposed neural network architecture is presented.  
\subsection{Neural Network Architecture}
We propose a deep learning model based on a cGAN, leveraging the \texttt{Pix2Pix} framework~\cite{isola2017image}. The model consists of two main components: a generator and a discriminator. The generator follows an encoder-decoder architecture for feature extraction and reconstruction, while the discriminator classifies the generator's output as real or fake. Convolutional layers handle feature extraction, and fully connected layers process abstract information. The proposed generator follows a classic U-Net architecture, and the discriminator is a Markovian PatchGAN classifier~\cite{isola2017image}. We describe each component in brief  for the sake of completeness below.
\paragraph{Generator Network}:
The generator converts input images into output images using an encoder-decoder architecture. The encoder extracts features while the decoder reconstructs the target image. Skip connections are used between encoder and decoder layers to preserve spatial details.

\paragraph{Encoder:} 
The encoder processes input images of size $(256 \times 256 \times 1)$, downsampling via convolutional layers, batch normalization, and Leaky ReLU activation. This results in a compressed representation of the input image. 
Convolutional layers progressively reduce the spatial dimensions with strides of $(2 \times 2)$, e.g., from $(256 \times 256 \times 1)$ to $(128 \times 128 \times 64)$~\cite{chadha2024pytoacnn}. Batch normalization stabilizes training, and Leaky ReLU maintains gradient flow.

\paragraph{Adaptive Dense Layer:} 
At the bottleneck, the encoder's output is flattened and passed through dense layers to handle abstract features. One can adjust the number of neurons based on the complexity of the task.

\paragraph{Decoder:} 
The decoder upsamples the feature maps back to the original image size using transpose convolution layers. Skip connections from the encoder to the decoder to ensure spatial details are retained. ReLU activation and zero padding are applied during upsampling to maintain size consistency.

\paragraph{Discriminator Network:} 
The discriminator classifies image pairs (input and generated output) as real or fake using convolutional layers. A PatchGAN approach classifies individual image patches~\cite{isola2017image}, improving results on finer details. Leaky ReLU is used for stability, and Sigmoid activation at the output provides the probability of the image being real.

This architecture efficiently combines convolutional layers for feature extraction with dense layers for abstract processing, and the cGAN framework ensures that generated images are conditioned on input data, making it well-suited for tasks like image-to-image translation.

\begin{figure}[h]
	\centering
	\begin{subfigure}[t]{\textwidth}
		\centering
		\includegraphics[scale=0.45]{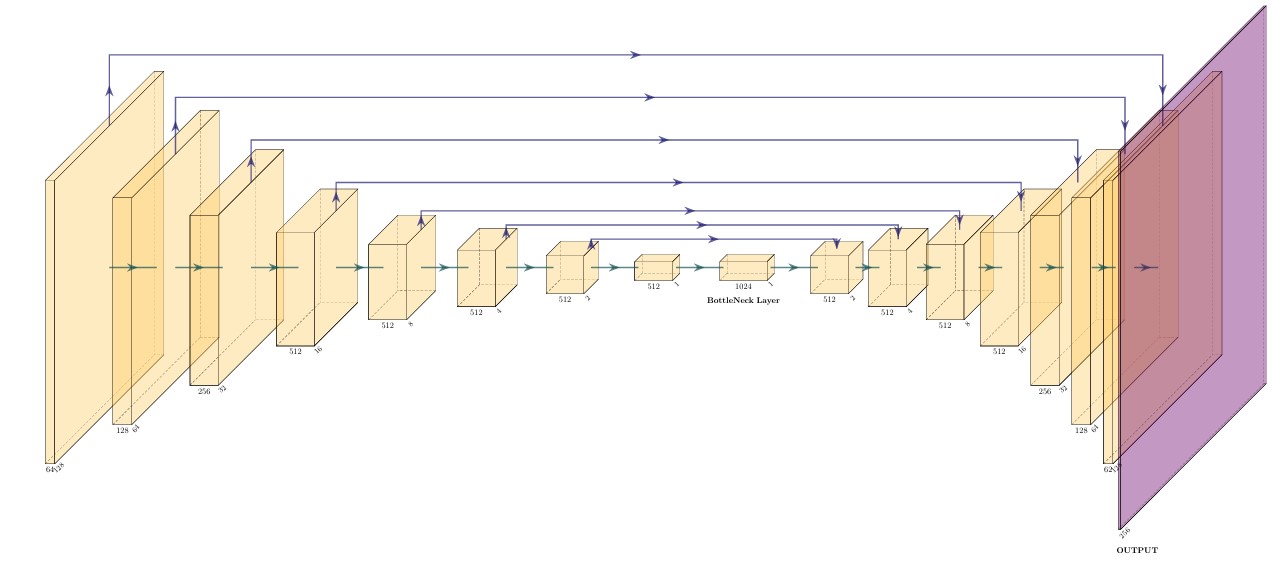}
		\caption{Proposed Generator Architecture}
	\end{subfigure}
	\begin{subfigure}[t]{\textwidth}
		\centering
		\includegraphics[scale=0.325]{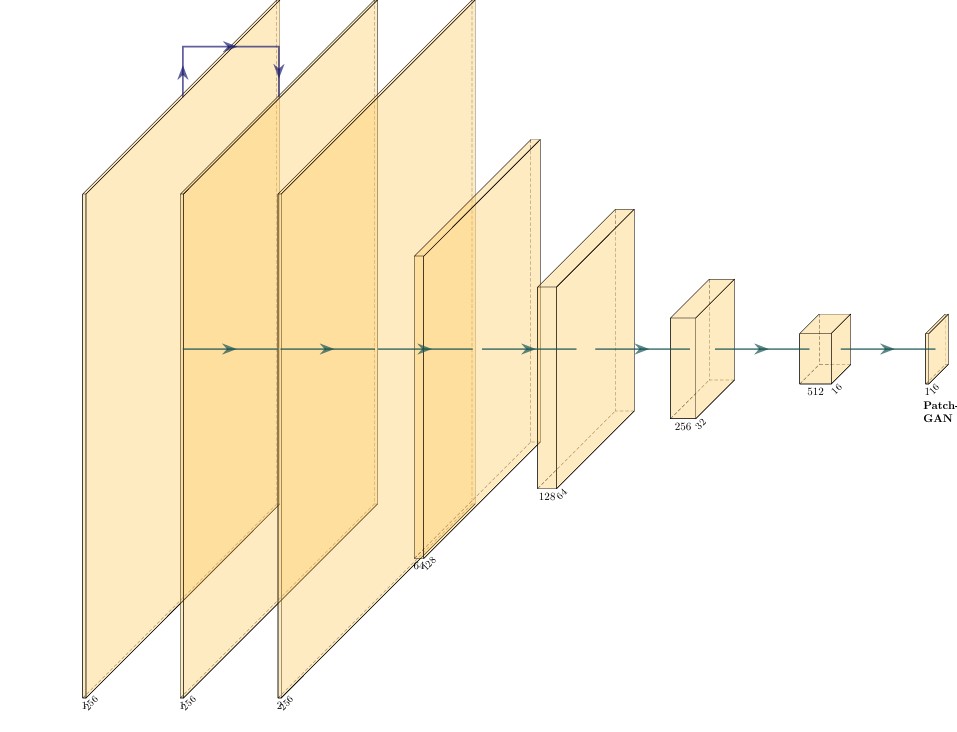}
		\caption{Proposed Discriminator Architecture}
	\end{subfigure}
	\caption{Different parts of the proposed architectures}
\end{figure}

\subsection{Input Mechanism}
We employ a novel input mechanism for scalar conditions~\cite{chadha2024pytoacnn} in the proposed conditional GAN model~\cite{isola2017image}. Scalar values are normalized within a predefined range \([min, max]\) and transformed into black-and-white images. The number of black-to-white pixels in the image corresponds to the scalar's value post-normalization, thus encoding the magnitude of  condition as a visual representation. This technique allows scalar conditions to be seamlessly incorporated into the generator and discriminator networks, facilitating condition-based variations while maintaining the convolutional nature of the model~\cite{chadha2024pytoacnn}. 

We denote the generator and discriminator using $G$ and $D$ letters, respectively. GANs learn a mapping form random noise vector $z$ to output image $y$, $G:z\to y$~\cite{goodfellow2014generative}. Whereas, cGANs learn a mapping from observed image $x$ and random noise $z$, to $y$, $G:{x,\,z}\to y$. Readers can refer to~~\cite{goodfellow2014generative,chadha2024pytoacnn,isola2017image} for more details. 

\subsection{Loss Function and Optimization}
The general GAN loss (BCE loss, which is used for binary classification) is written as~\cite{goodfellow2014generative}: \\
\begin{equation}
	\label{eq:minimaxgame-definition}
	\min_G \max_D V(G, D) = \mathbb{E}_{{x} \sim p_{\text{data}}({x})}[\log D({x})] + \mathbb{E}_{{z} \sim p_{{z}}({z})}[\log (1 - D(G({z})))],
\end{equation}
where $D(x):$ discriminator's prediction that input $x$ is real, $G(z):$ generator's output for random input $z$, $p_{\text{data}}({x})$: true data distribution, and $p_{{z}}({z})$: prior noise distribution (e.g., Gaussian or Uniform). $\mathbb{E}$ represents expectation operator, averaging all possible samples from a distribution~\cite{goodfellow2014generative}.

Likewise, for the cGAN, one writes the loss function as~\cite{mirza2014conditional}
\begin{align}
	\mathcal{L}_{cGAN}(G,D) = \mathbb{E}_{x,y}[\log D(x,y)] + \mathbb{E}_{x,z}[\log (1-D(x,G(x,z))],\label{cGAN_equation}
\end{align}
where $G$ tries to minimize this objective against an adversarial $D$ that tries to maximize it, i.e.
\begin{align}
	G^{*}  = \arg\min_G \max_D \mathcal{L}_{cGAN}(G,D).
\end{align}

Previous approaches have demonstrated that combining the GAN objective with a more conventional loss function, such as the L1 distance, can improve its performance. The discriminator's role does not changed. The generator is designed not only to deceive the discriminator but also to closely approximate the ground truth output in terms  an $L_2$ sense. We explore the mentioned option, i.e., we use $L_1$ distance (MAE loss) rather than $L_2$ as $L_1$ encourages less blurring as
\begin{align}
	\mathcal{L}_{L1}(G) = \mathbb{E}_{x,y,z}[\lVert{y-G(x,z)}_1\rVert].\label{L1_equation}
\end{align}
The final objective of such a \texttt{Pix2Pix} framework-based GAN is given as~\cite{isola2017image}:
\begin{align}
	G^*  = \arg\min_G\max_D \mathcal{L}_{cGAN}(G,D) + \lambda \mathcal{L}_{L1}(G).\label{full_objective}
\end{align}
which is the BCE loss with a weighted MAE loss to achieve both adversarial learning and accurate design reconstruction. In this setup, the discriminator's role is to distinguish real from generated designs, while the generator is tasked with two primary objectives: minimizing the BCE loss to fool the discriminator and generating outputs structurally consistent with the real images. To enhance reconstruction quality, the generator also minimizes the L1 distance (weighted by a factor, $\lambda = 100$) as the MAE loss, preferred over L2 for its ability to produce sharper outputs by reducing image blurring.
\begin{algorithm}[h!]
	\caption{Training process for the model}\label{alg:example}
	The ability of the model to capture intricate geometries and to avoid mode collapse is observed through this dynamic batch gradient descent-based training loop. The number of steps to apply to the discriminator, $k$, is a hyperparameter. We use $k=1$, the least expensive option, in our model.
	
	\textbf{for} a dynamic number of training iterations \textbf{for} its respective batch size \{1,2,4,8...\} which changes dynamically \textbf{do},
	\begin{itemize}
		\item \textbf{for} $k$ steps \textbf{do}
		\begin{itemize}
			\item Sample minibatch of $n$ examples, $\{ x^{(1)}, \dots, x^{(n)} \}$, along with their corresponding scalar conditions from the data-generating distribution $p_{data}(x)$, where  $n$ is the current batch size.
			\item Update D by ascending its stochastic gradient:
			\[
			\nabla_{\theta_d} \frac{1}{n} \sum_{i=1}^{n} \left[ \log D(x^{(i)}) + \log \left( 1 - D(G(z^{(i)})) \right) \right].
			\]
		\end{itemize}
		\item \textbf{end for}
		\item Sample minibatch of $n$ generated outputs $\{ z^{(1)}, \dots, z^{(n)} \}$ from prior $p_g(z)$.
		\item Update G by descending its stochastic gradient:
		\[
		\nabla_{\theta_g} \frac{1}{n} \sum_{i=1}^{n} \left[ \log \left( 1 - D(G(z^{(i)})) \right) + \lambda \cdot \text{MAE}(x^{(i)}, G(z^{(i)})) \right].
		\]
	\end{itemize}
	\textbf{end for}
\end{algorithm}

The generator defines a probability distribution G(z) over the training data, where z represents the latent input conditioned by scalar design parameters. The training loop then tunes the GAN's parameters, aligning the generator's distribution to match the dataset's underlying probability distribution. This adversarial training, regulated by the discriminator, ensures the generated designs adhere closely to the input specifications while maintaining structural fidelity, enabling the model to produce high-quality, optimized designs that meet precise performance constraints.

An algorithm follows (Algorithm~\ref{alg:example}), detailing the step-by-step workflow of the model, illustrating how the GAN iteratively optimizes the generator to produce high-quality, performance-specific designs. The corresponding algorithm is provided next.

\section{Results and Discussion} \label{Sec:3}
This section provides results for the cantilever beam and chair designs. We also report compliance and volume fraction errors for the optimized cantilever beam as per the method provided in~\cite{chadha2024pytoacnn}. 
\subsection{Topology optimization of a cantilever beam}
To determine the optimized designs with constant loads, typically, the following standard compliance minimization problem is solved with a given resource constraint \cite{kumar2023honeytop90}:
\begin{equation} \label{EQ:OPTI} 
	\begin{rcases}
		\begin{split}
			&{\min_{\bm{\rho}}} \quad C({\bm{\rho}}) = \mathbf{u}^\top  \mathbf{K}(\bm{\rho}) \mathbf{u} \\
			&\text{subjected to:}\\
			&\mathbf{K} \mathbf{u} -  \mathbf{F} = \mathbf{0}\\
			& V(\bm{\rho})-V^* \le 0\\
			&\quad\,\,\,\, \bm{0} \leq \bm{\rho} \leq \bm{1} 
		\end{split}
	\end{rcases},
\end{equation} 
where $C$ indicates compliance of the structure. $\mathbf{u}$ and $\mathbf{K}$ denote the global displacement vector and stiffness matrix, respectively. $V$ and $V^*$ indicate the current and permitted volume of the design domain, respectively.

We test the proposed neural network model to obtain the optimized design for a cantilever beam, and 59 such MATLAB-generated designs are used as the training dataset. The volume fraction and Possion's ratio are taken as the input for the model. 

The results generated by the network and MATLAB code~\cite{andreassen2011efficient} are depicted in Table~\ref{T:T1} side-by-side. The network provides the output image in grayscale and from that we determine  $\texttt{xphys}_\text{GAN}$. $V_\text{GAN} = \text{mean}(\texttt{xphys}_\text{GAN})\times nel$ is calculated~\cite{chadha2024pytoacnn}; thus, $V_\text{err}=\frac{V^* - V_\text{GAN}}{V_f}\times 100 \%$. Additionally, $\texttt{xphys}_\text{GAN}$ is used to determine the corresponding compliance $C_\text{GAN}$; and thus, $C_\text{err} = \frac{C_\text{act} - C_\text{GAN}}{C_\text{GAN}}\times 100 \%$, where $C_\text{act}$ is directly obtained from the MATLAB code. 
\begin{table}[h]
	\centering
	
	\begin{tabular}{|c|c|c|c|c|}
		\hline
		\textbf{Input} & \textbf{Ground Truth} & \textbf{Generated Image} & {$V_\text{err}(\%)$} & $C_\text{err}(\%)$ \\ 
		\hline
		$V_f=0.25, \nu=0.4$ & \includegraphics[scale=0.14]{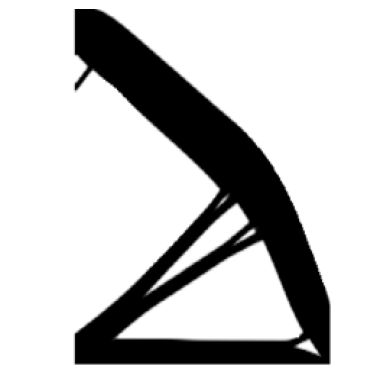} & \includegraphics[scale=0.15]{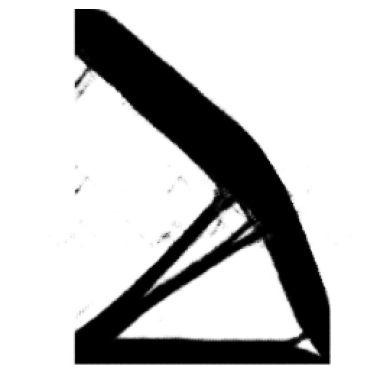} &{0.02047} & {1.79506} \\ 
		\hline
		$V_f=0.35,\nu=0.5$ & \includegraphics[scale=0.14]{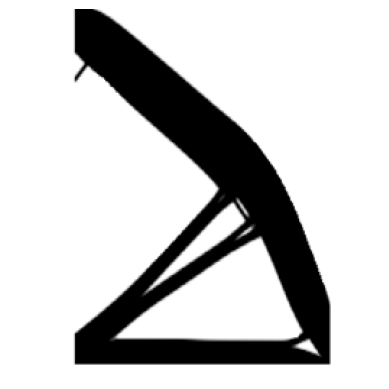} & \includegraphics[scale=0.15]{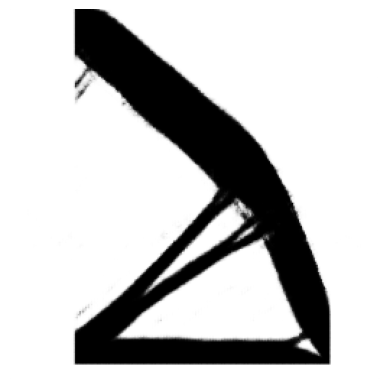} & {0.01912} & {1.5560} \\
		\hline
		$V_f=0.4, \nu=0.3$ & \includegraphics[scale=0.14]{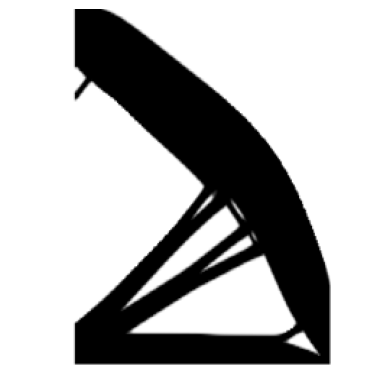} & \includegraphics[scale=0.15]{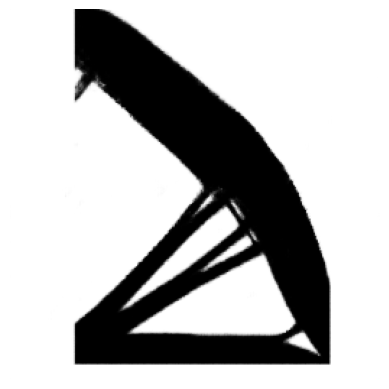} & {0.02508} & {1.48815} \\
		\hline
		$V_f=0.45, \nu=0.3$ & \includegraphics[scale=0.14]{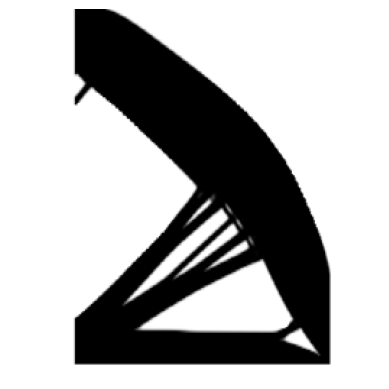} & \includegraphics[scale=0.15]{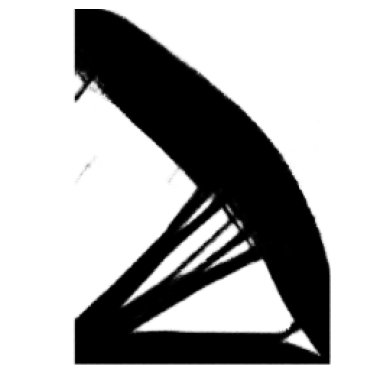} & {0.03391} & {1.67573} \\
		\hline
		$V_f=0.55,\nu=0.4$ & \includegraphics[scale=0.14]{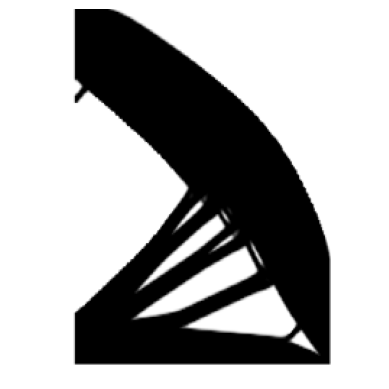} & \includegraphics[scale=0.15]{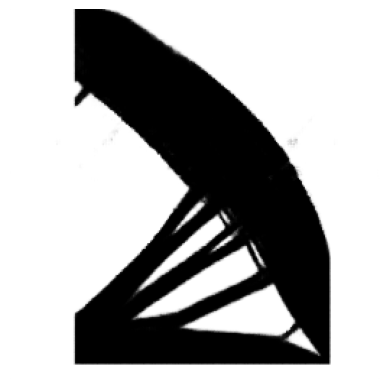} & {0.02959} & {0.44609} \\
		\hline
	\end{tabular} \\
	\caption{Results of the cantilever beam problem. ($V_f$ - Volume fraction, $\nu$ - Poisson's ratio)}\label{T:T1}
\end{table}

One notes that $V_\text{err}(\%)$ and $C_\text{err}(\%)$~(Table~\ref{T:T1}) are insignificant, indicating the success and robustness of the proposed  GAN model. The model provides optimized designs in a fraction of a second, reducing computational time for generating the optimized designs once it is trained for. Future avenues for the model include using it to solve different TO problems involving multi-physics. 
\begin{table}[h]
	\centering
	\begin{tabular}{|c|*{10}{c|}}
		\hline
		\textbf{Yaw Angle} & \multicolumn{10}{|c|}{\textbf{Chair Configurations}} \\
		\hline
		$0^o$ & \includegraphics[scale=0.1]{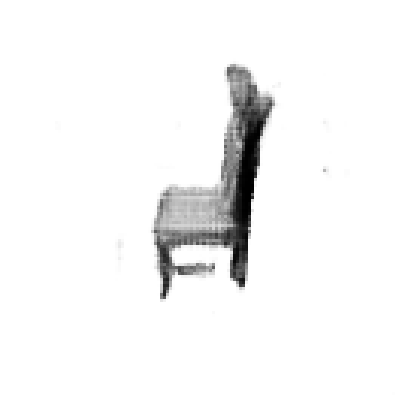} & 
		\includegraphics[scale=0.1]{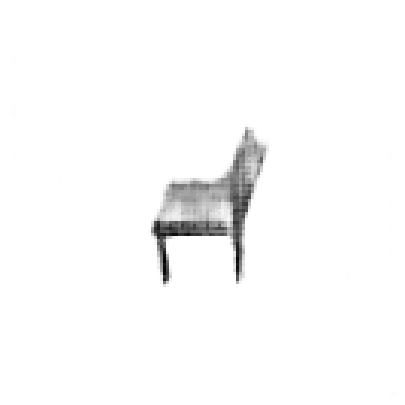} & 
		\includegraphics[scale=0.1]{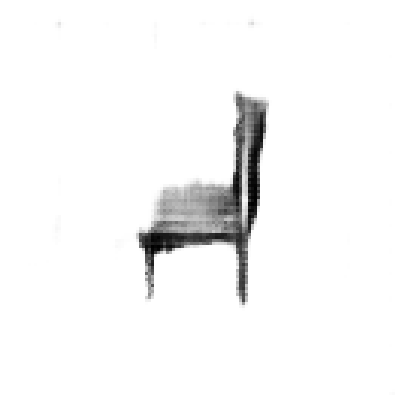} & 
		\includegraphics[scale=0.1]{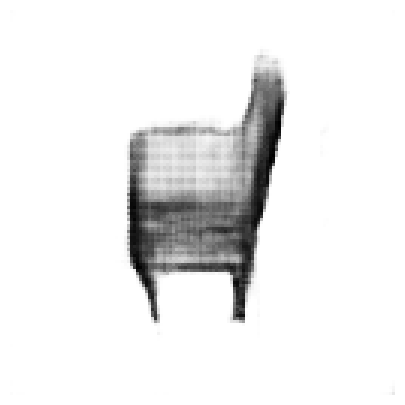} & 
		\includegraphics[scale=0.1]{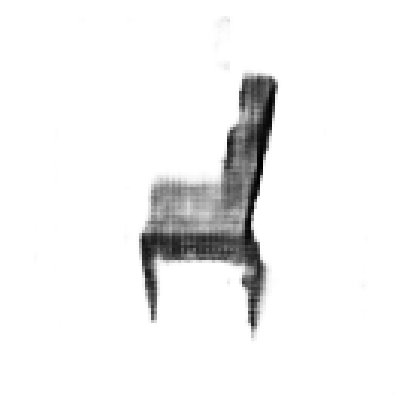} & 
		\includegraphics[scale=0.1]{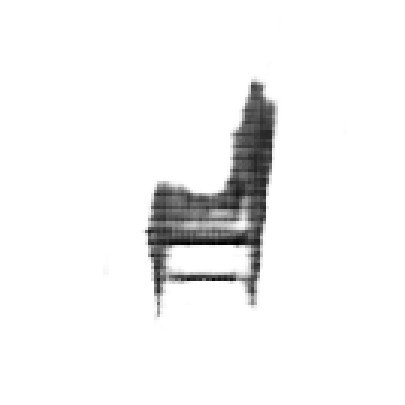} &
		\includegraphics[scale=0.1]{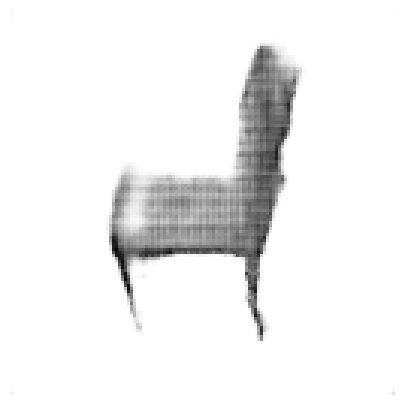} & 
		\includegraphics[scale=0.1]{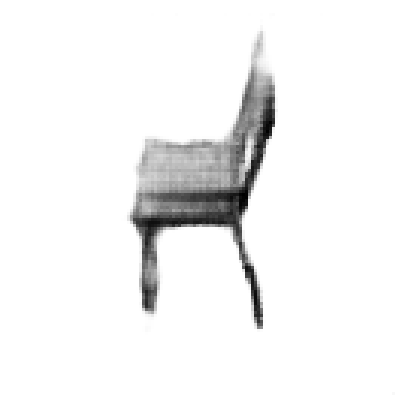} & 
		\includegraphics[scale=0.1]{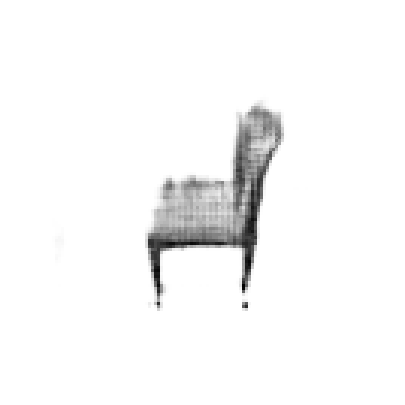} & 
		\includegraphics[scale=0.1]{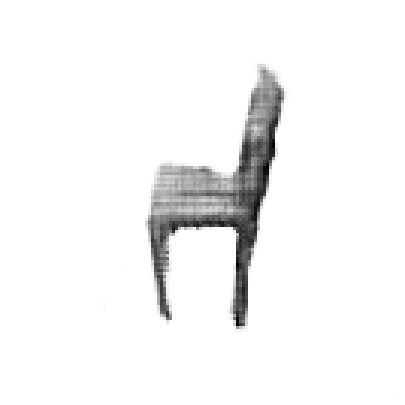} \\
		\hline
		$5.4^o$ & \includegraphics[scale=0.1]{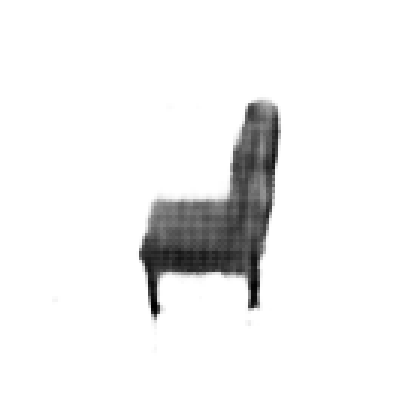} &
		\includegraphics[scale=0.1]{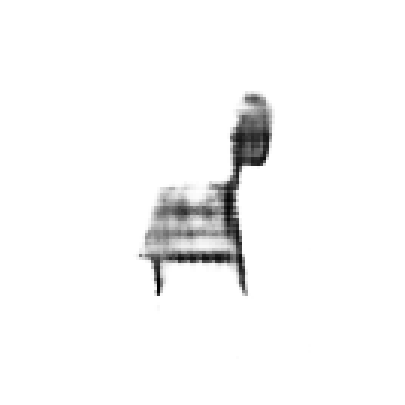} & 
		\includegraphics[scale=0.1]{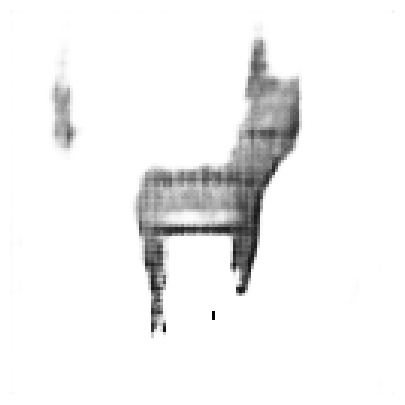} & 
		\includegraphics[scale=0.1]{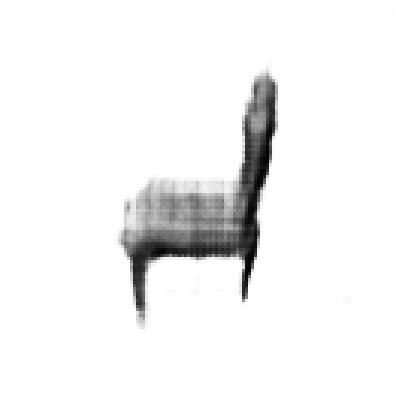} & 
		\includegraphics[scale=0.1]{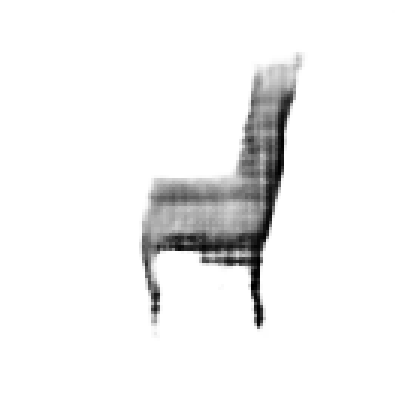} & 
		\includegraphics[scale=0.1]{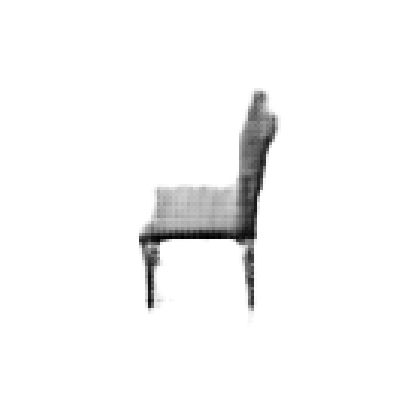} & 
		\includegraphics[scale=0.1]{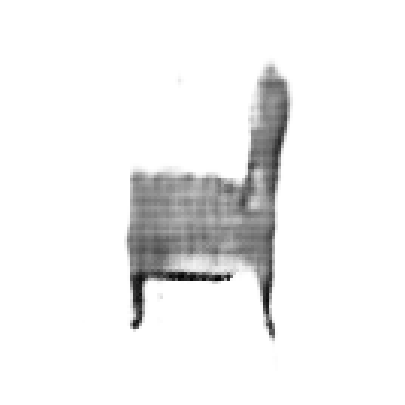} & 
		\includegraphics[scale=0.1]{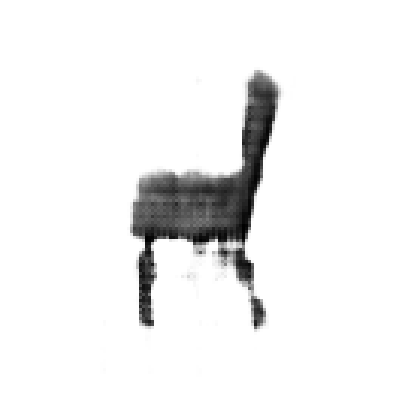} & 
		\includegraphics[scale=0.1]{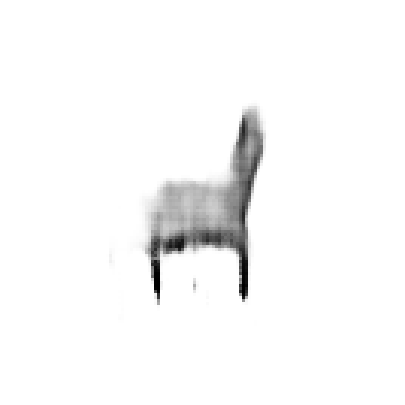} &
		\includegraphics[scale=0.1]{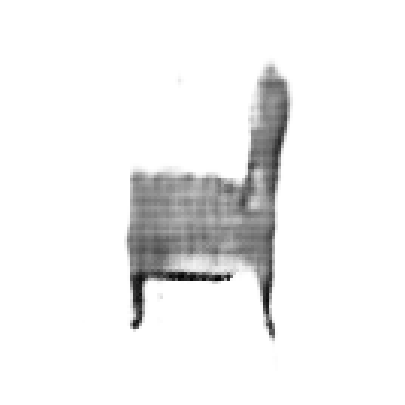} \\
		\hline
		$10.2^o$ & \includegraphics[scale=0.1]{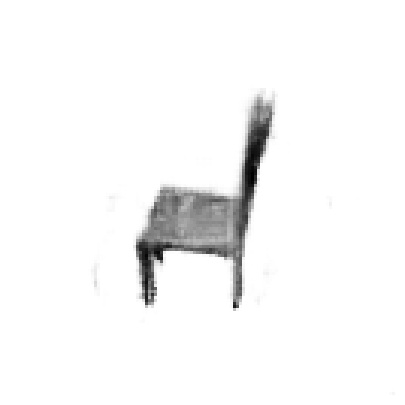} &
		\includegraphics[scale=0.1]{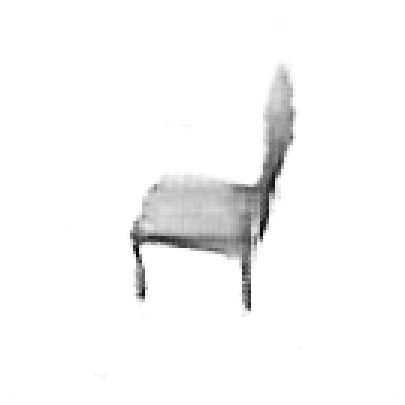} & 
		\includegraphics[scale=0.1]{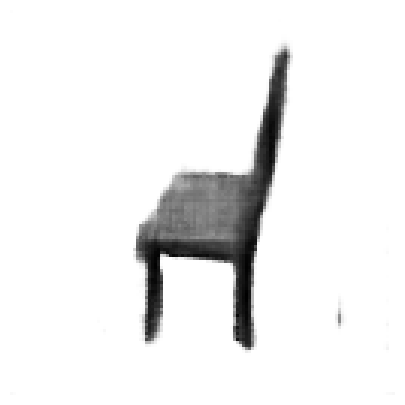} & 
		\includegraphics[scale=0.1]{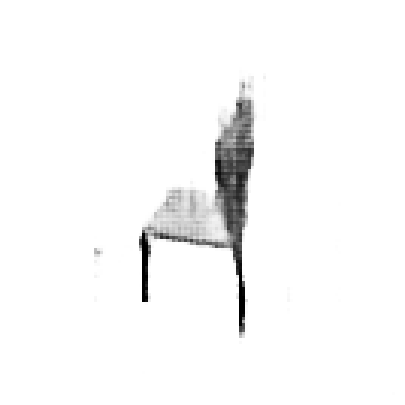} & 
		\includegraphics[scale=0.1]{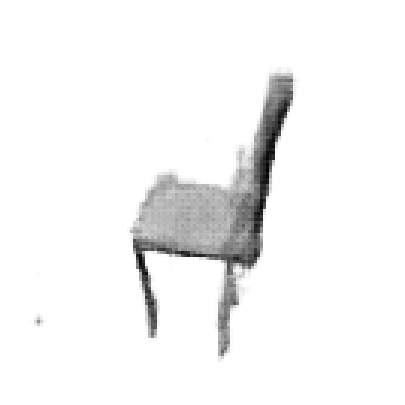} & 
		\includegraphics[scale=0.1]{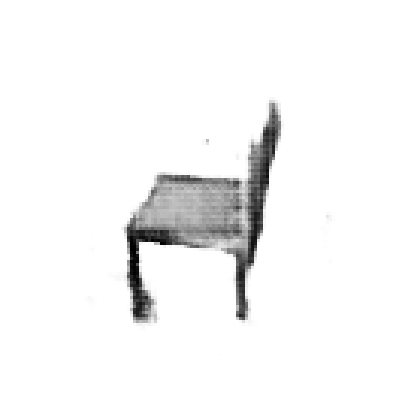} & 
		\includegraphics[scale=0.1]{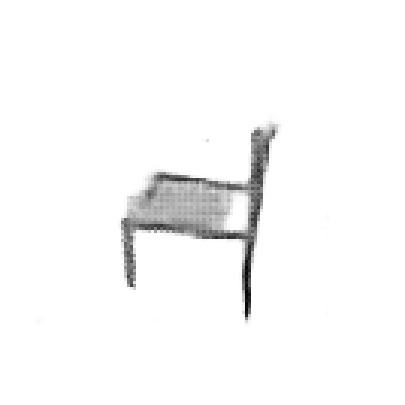} & 
		\includegraphics[scale=0.1]{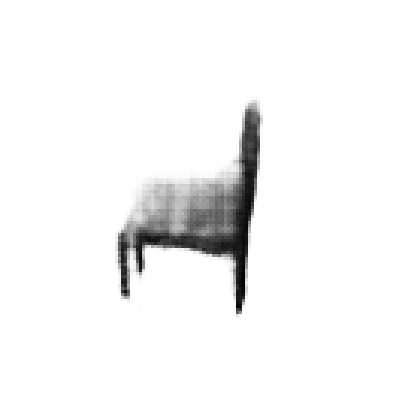} & 
		\includegraphics[scale=0.1]{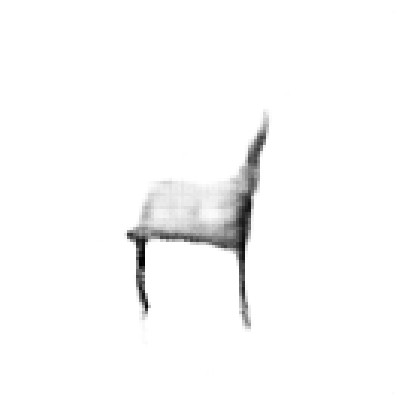} & 
		\includegraphics[scale=0.1]{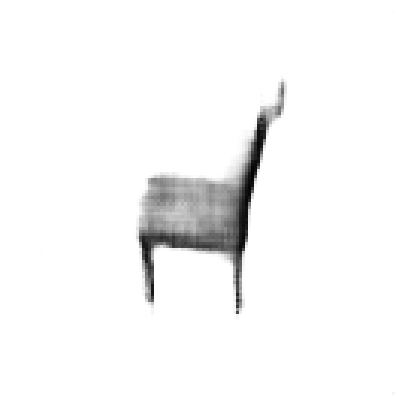} \\
		\hline
		$14.9^o$ & \includegraphics[scale=0.1]{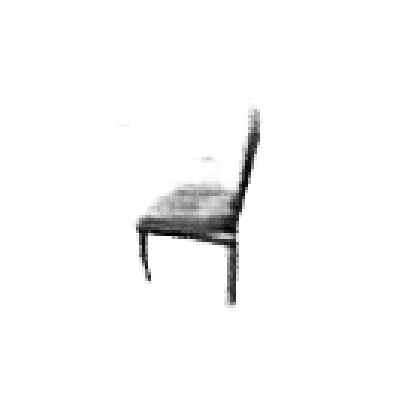} & 
		\includegraphics[scale=0.1]{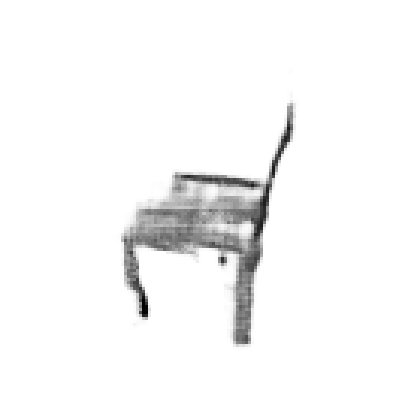} & 
		\includegraphics[scale=0.1]{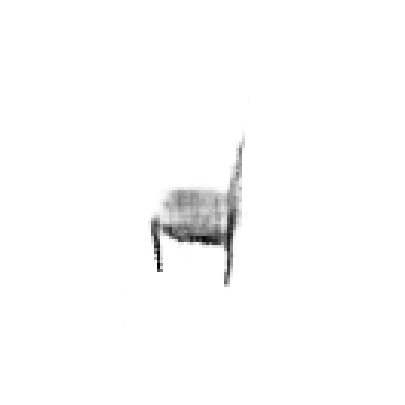} & 
		\includegraphics[scale=0.1]{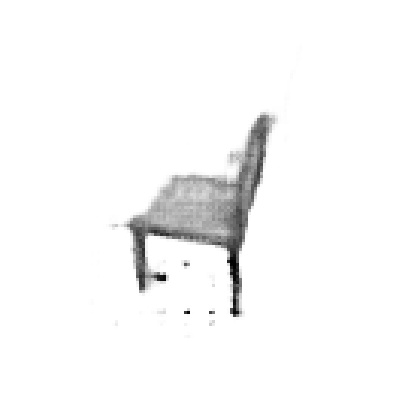} & 
		\includegraphics[scale=0.1]{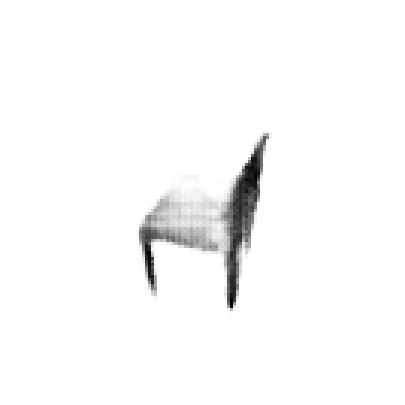} &
		\includegraphics[scale=0.1]{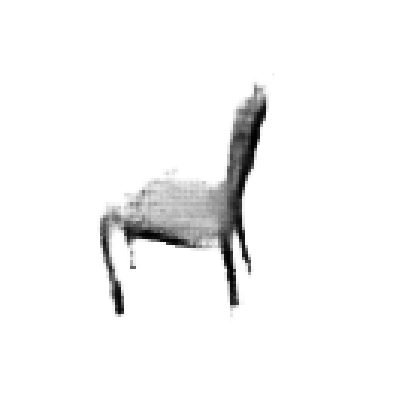} &
		\includegraphics[scale=0.1]{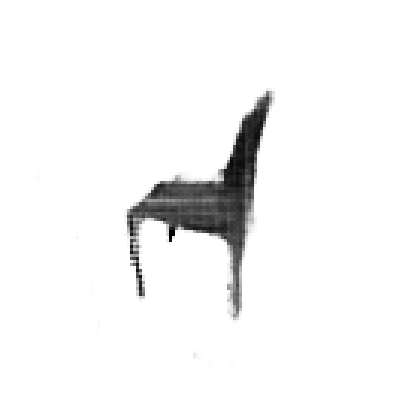} &
		\includegraphics[scale=0.1]{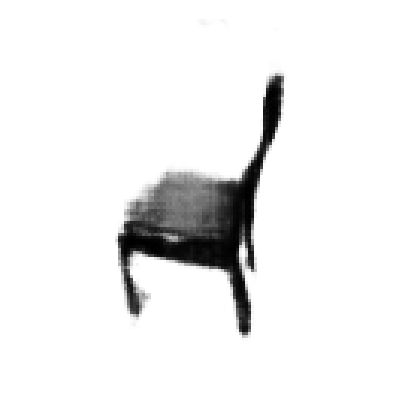} & 
		\includegraphics[scale=0.1]{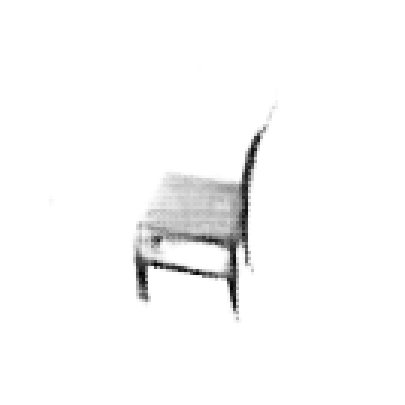} & 
		\includegraphics[scale=0.1]{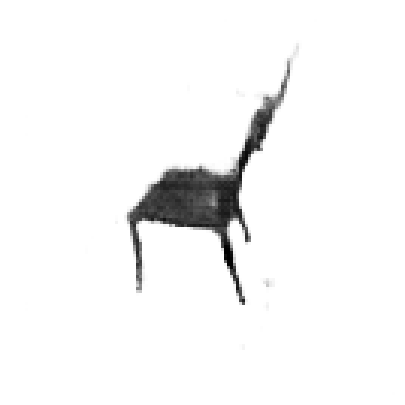} \\
		\hline
		$20^o$ & \includegraphics[scale=0.1]{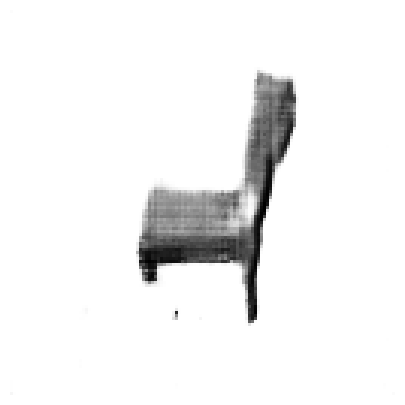} & 
		\includegraphics[scale=0.1]{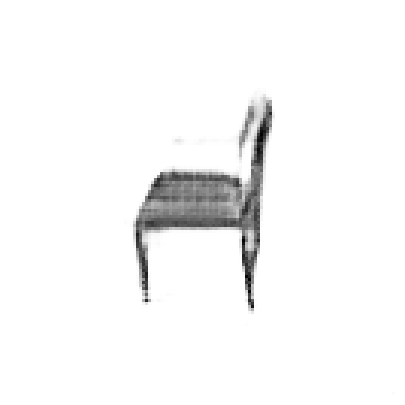} & 
		\includegraphics[scale=0.1]{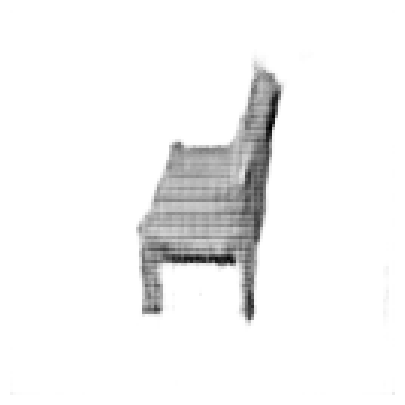} & 
		\includegraphics[scale=0.1]{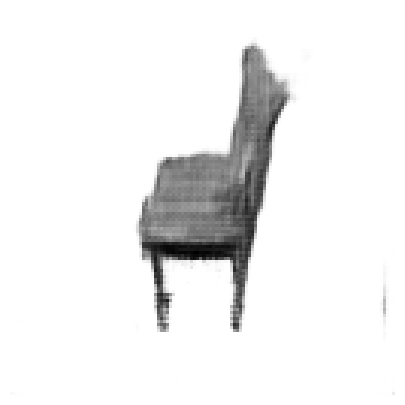} & 
		\includegraphics[scale=0.1]{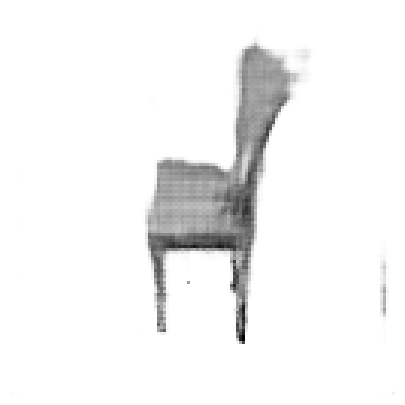} & 
		\includegraphics[scale=0.1]{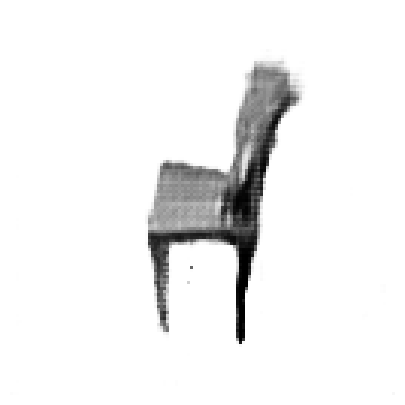} & 
		\includegraphics[scale=0.1]{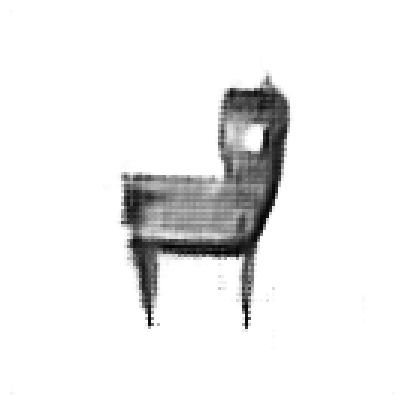} & 
		\includegraphics[scale=0.1]{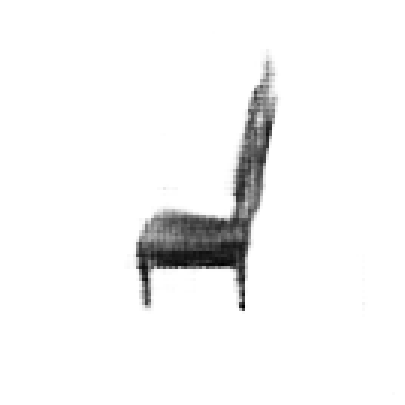} & 
		\includegraphics[scale=0.1]{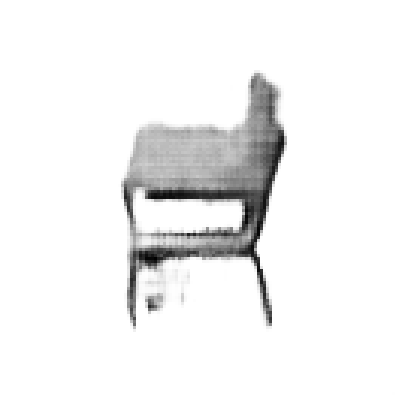} & 
		\includegraphics[scale=0.1]{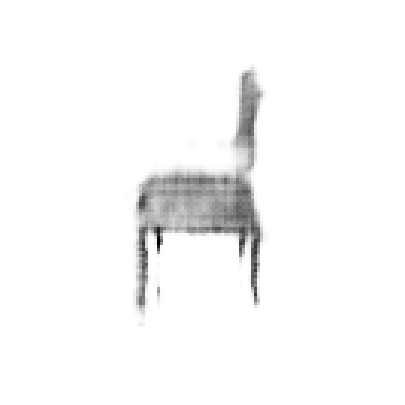} \\
		\hline
		$23.8^o$ & \includegraphics[scale=0.1]{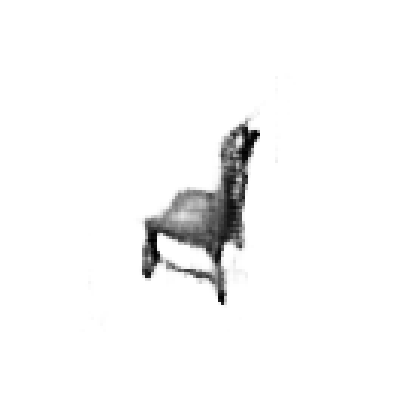} & 
		\includegraphics[scale=0.1]{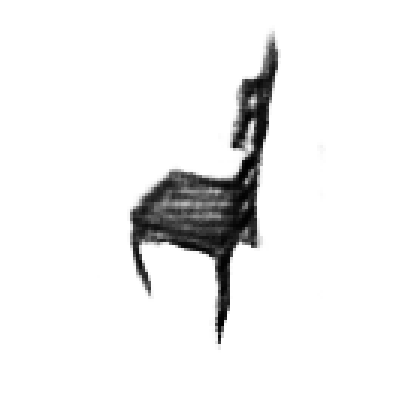} & 
		\includegraphics[scale=0.1]{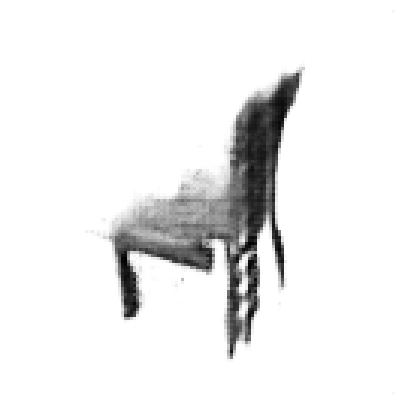} & 
		\includegraphics[scale=0.1]{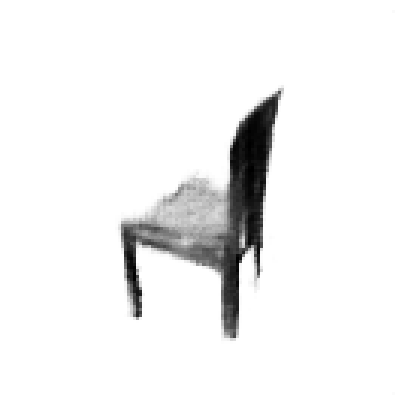} & 
		\includegraphics[scale=0.1]{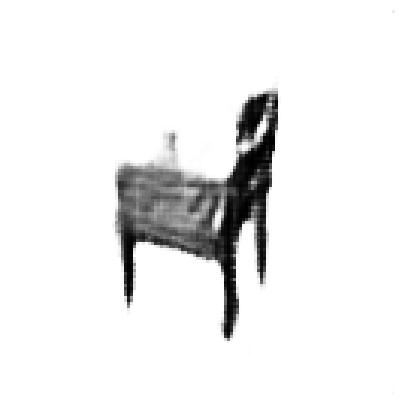} & 
		\includegraphics[scale=0.1]{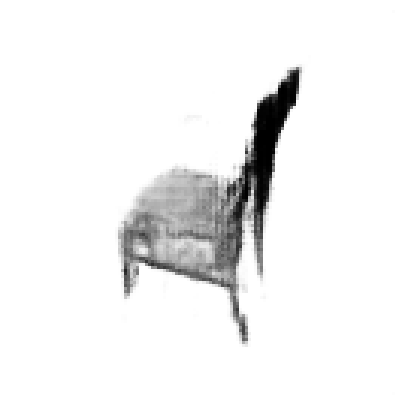} & 
		\includegraphics[scale=0.1]{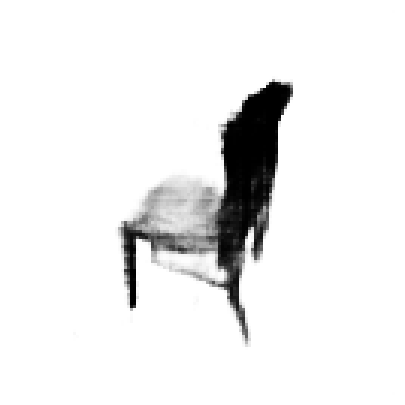} & 
		\includegraphics[scale=0.1]{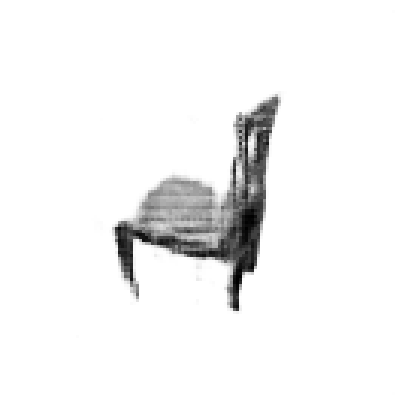} & 
		\includegraphics[scale=0.1]{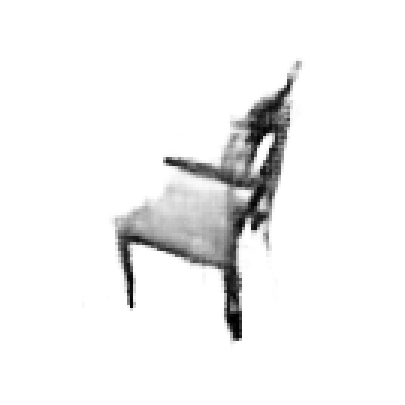} & 
		\includegraphics[scale=0.1]{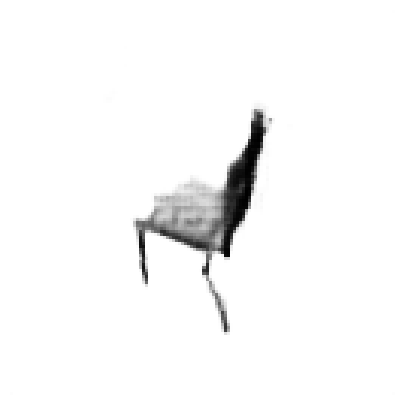} \\
		\hline
		$59^o$ & \includegraphics[scale=0.1]{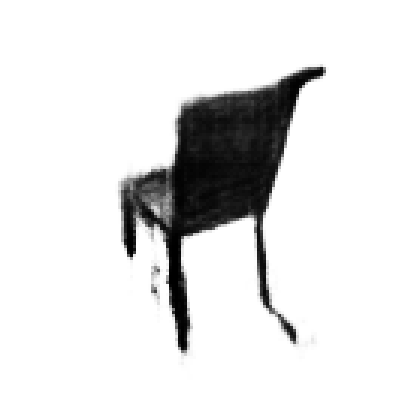} &
		\includegraphics[scale=0.1]{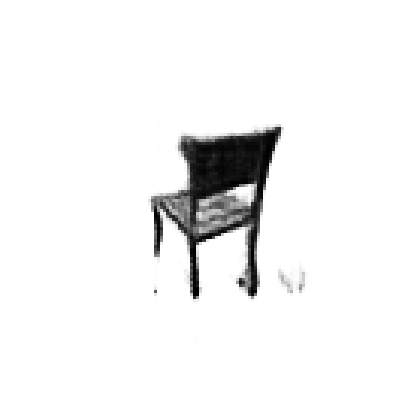} & 
		\includegraphics[scale=0.1]{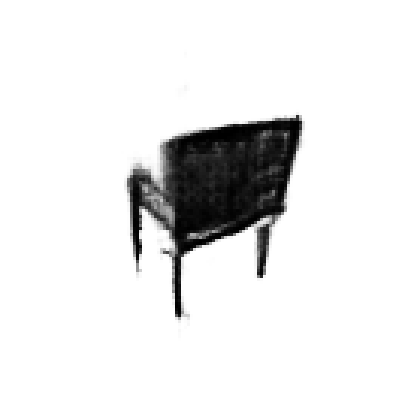} & 
		\includegraphics[scale=0.1]{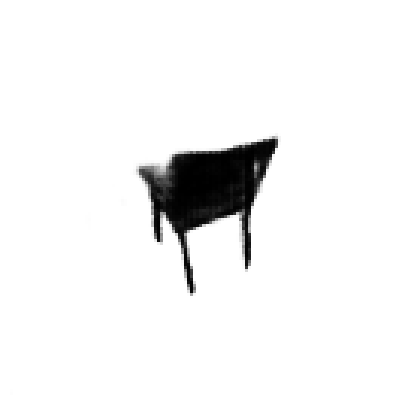} & 
		\includegraphics[scale=0.1]{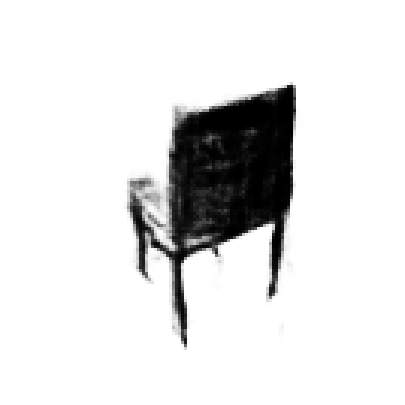} & 
		\includegraphics[scale=0.1]{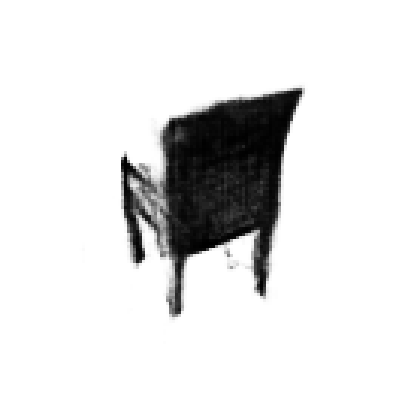} & 
		\includegraphics[scale=0.1]{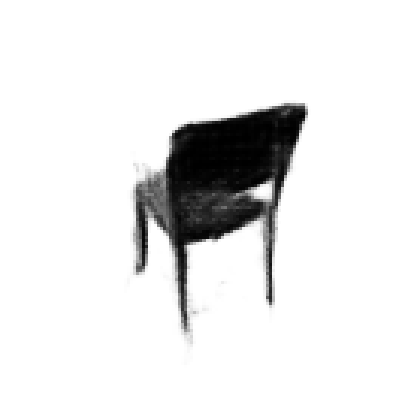} & 
		\includegraphics[scale=0.1]{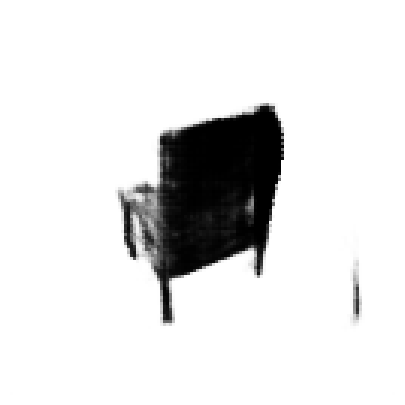} & 
		\includegraphics[scale=0.1]{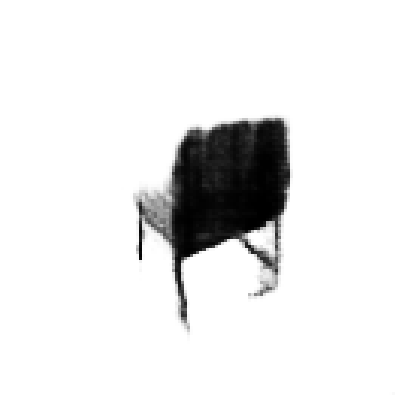} & 
		\includegraphics[scale=0.1]{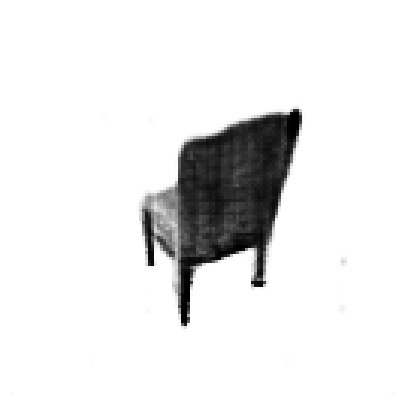} \\
		\hline
		$68^o$ & \includegraphics[scale=0.1]{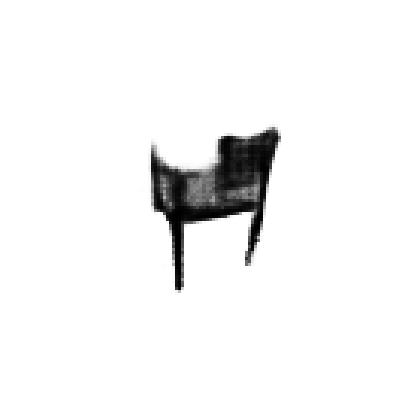} & 
		\includegraphics[scale=0.1]{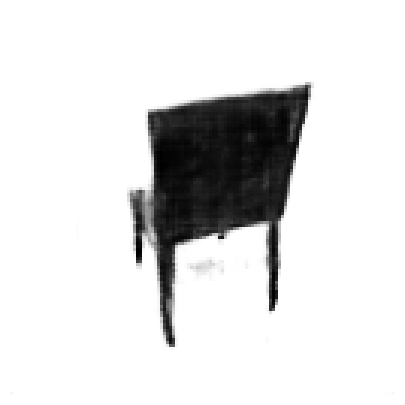} & 
		\includegraphics[scale=0.1]{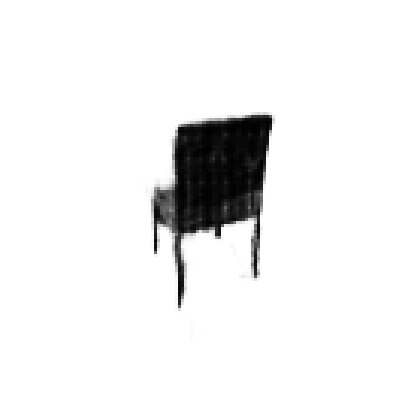} & 
		\includegraphics[scale=0.1]{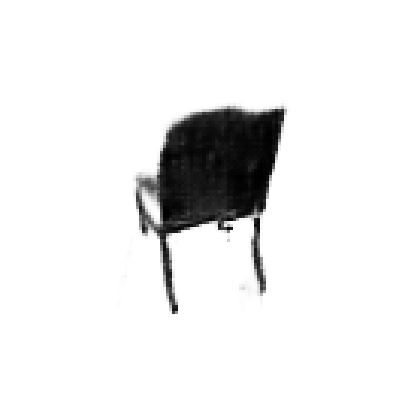} & 
		\includegraphics[scale=0.1]{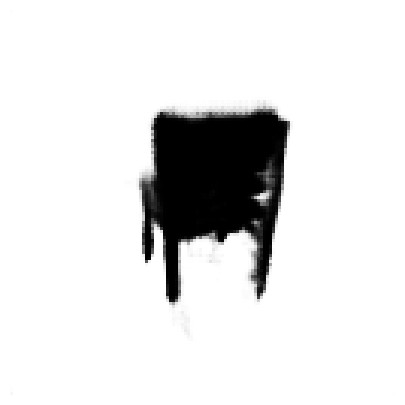} & 
		\includegraphics[scale=0.1]{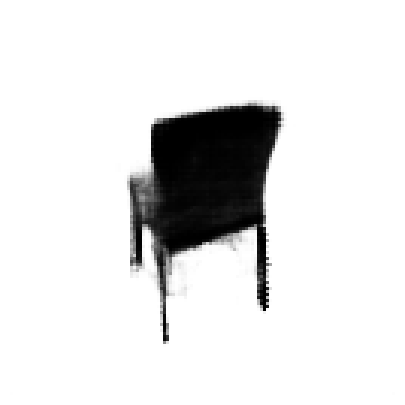} & 
		\includegraphics[scale=0.1]{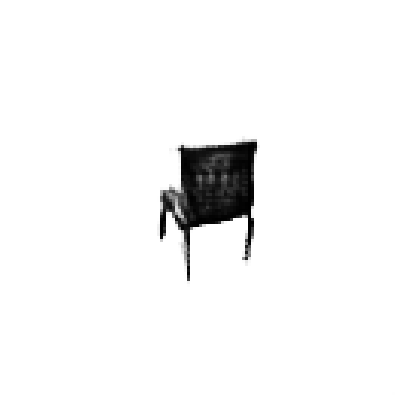} & 
		\includegraphics[scale=0.1]{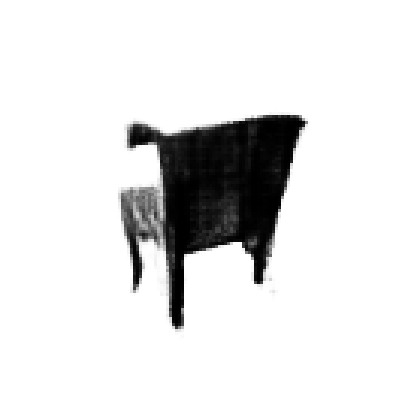} & 
		\includegraphics[scale=0.1]{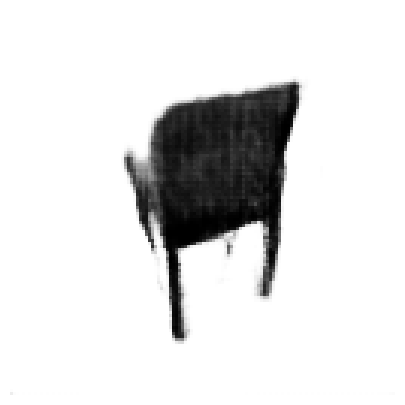} & 
		\includegraphics[scale=0.1]{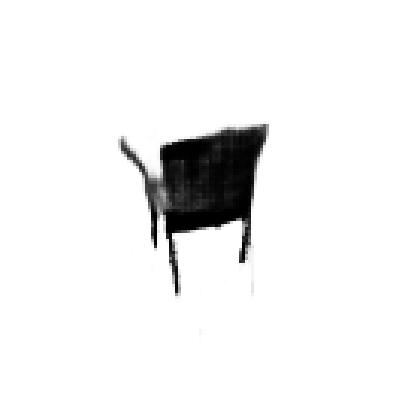} \\ 
		\hline
		$85^o$ & \includegraphics[scale=0.1]{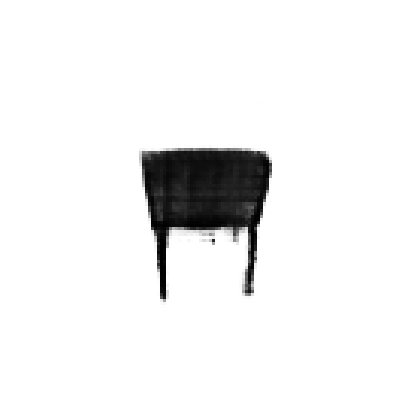} &
		\includegraphics[scale=0.1]{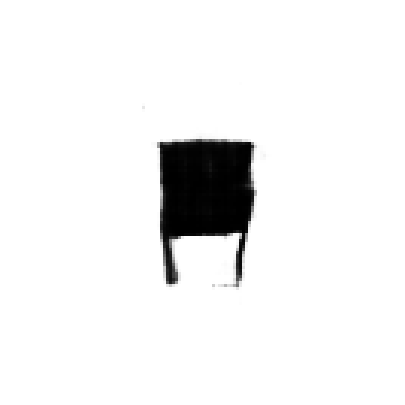} & 
		\includegraphics[scale=0.1]{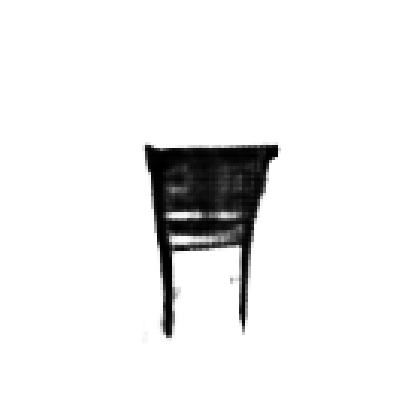} & 
		\includegraphics[scale=0.1]{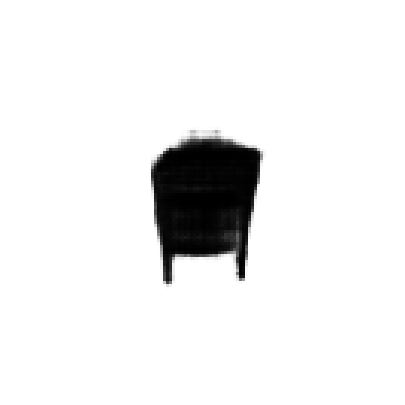} & 
		\includegraphics[scale=0.1]{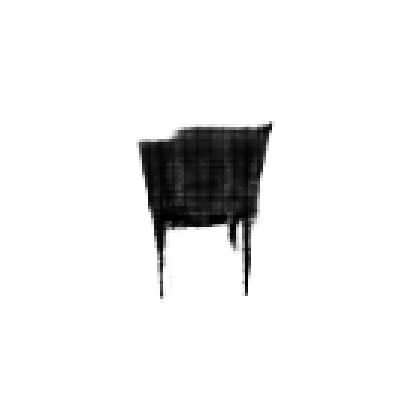} & 
		\includegraphics[scale=0.1]{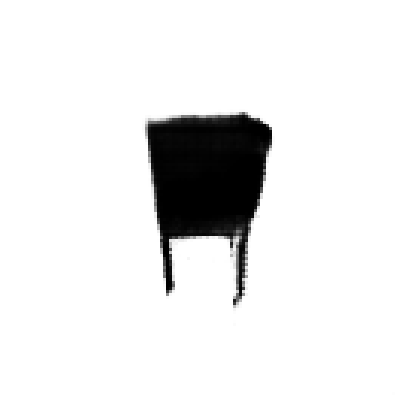} & 
		\includegraphics[scale=0.1]{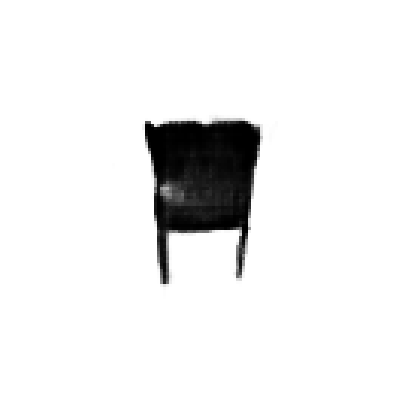} & 
		\includegraphics[scale=0.1]{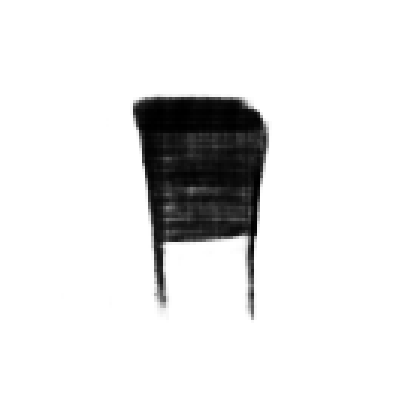} & 
		\includegraphics[scale=0.1]{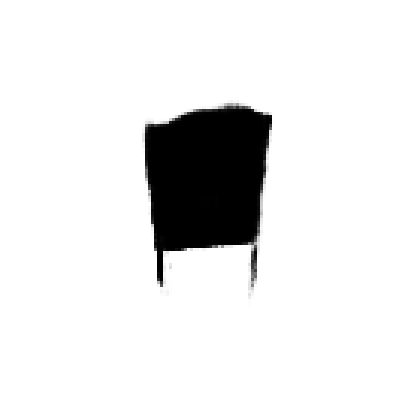} & 
		\includegraphics[scale=0.1]{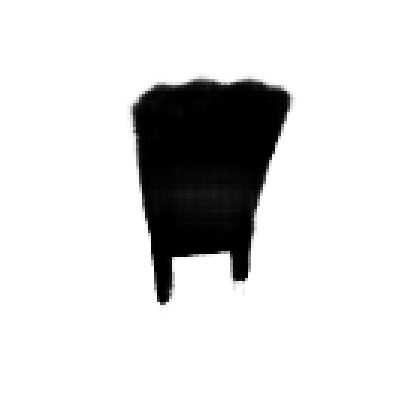} \\
		\hline
	\end{tabular}
	\caption{Results of chair design prediction problem}\label{T:T2}
\end{table}
\subsection{Chair Design Prediction}
For this problem, we train the model to predict different chair configurations at a given yaw angle. The training dataset consists of around 2,65,000 images. 

Table~\ref{T:T2} presents the results for various chair designs. For each Yaw angle, the model predicts 10 different chair designs. Users can select a design based on their specific requirements and comfort preferences.

The proposed GAN-based approach for cantilever beam topology optimization and chair design optimization demonstrates significant potential in generating optimal geometries from the provided dataset. This method offers a scalable and efficient solution, particularly for applications requiring rapid design iterations. However, there remains room for improvement. Enhancing the model's sensitivity to specific physical constraints, such as material properties and stress distribution, can result in more practical and robust designs.

Furthermore, by learning latent representations that enable the generation of diverse designs under the same constraints presents an exciting opportunity. Architectures like Variational Autoencoders (VAEs) can also be employed. Future directions include developing hybrid models combining VAEs with GANs and other generative techniques to enhance design flexibility and performance.
\section{Conclusions and Future directions} \label{Sec:4}
This work demonstrates the potential of Generative Adversarial Networks (GANs), specifically the proposed GO-GAN architecture, for efficient, and iterative design improvement in industrial applications. This is specifically illustrated here by example studies on cantilever beam and chair
design optimizations. GO-GAN effectively generates structures from user-specified parameters. However, model performance and generalization are currently limited by the size and diversity of the training dataset. Furthermore, while optimized within certain parameters, the current model
needs to explicitly incorporate complex physical constraints such as material properties and load-
bearing capacities. Like other GANs, GO-GAN's performance is sensitive to training
hyperparameters, requiring potentially time-consuming fine-tuning. Future research directions include integrating multiple optimal designs, increasing sensitivity to physical constraints,
expanding to multi-physics optimization objectives, and exploring alternative generative models
to further enhance engineering and manufacturing design practices by improving flexibility and
optimization efficiency.


\begin{thebibliography}{10}
	
	\bibitem{shin2023topologyoptimizationmachinelearning}
	S.~Shin, D.~Shin, and N.~Kang, ``Topology optimization via machine learning and
	deep learning: a review,'' {\em Journal of Computational Design and
		Engineering}, vol.~10, pp.~1736--1766, 07 2023.
	
	\bibitem{Abueidda_2020}
	D.~W. Abueidda, S.~Koric, and N.~A. Sobh, ``Topology optimization of 2d
	structures with nonlinearities using deep learning,'' {\em Computers \&
		Structures}, vol.~237, p.~106283, Sept. 2020.
	
	\bibitem{Rade_2021}
	J.~Rade, A.~Balu, E.~Herron, J.~Pathak, R.~Ranade, S.~Sarkar, and
	A.~Krishnamurthy, ``Algorithmically-consistent deep learning frameworks for
	structural topology optimization,'' {\em Engineering Applications of
		Artificial Intelligence}, vol.~106, p.~104483, Nov. 2021.
	
	\bibitem{goodfellow2014generative}
	I.~Goodfellow, J.~Pouget-Abadie, M.~Mirza, B.~Xu, D.~Warde-Farley, S.~Ozair,
	A.~Courville, and Y.~Bengio, ``Generative adversarial nets,'' {\em Advances
		in neural information processing systems}, vol.~27, 2014.
	
	\bibitem{app131910664}
	J.~Y. Seong, S.-m. Ji, D.-h. Choi, S.~Lee, and S.~Lee, ``Optimizing generative
	adversarial network (gan) models for non-pneumatic tire design,'' {\em
		Applied Sciences}, vol.~13, no.~19, 2023.
	
	\bibitem{ding2023continuousconditionalgenerativeadversarial}
	X.~Ding, Y.~Wang, Z.~Xu, W.~J. Welch, and Z.~J. Wang, ``Continuous conditional
	generative adversarial networks: Novel empirical losses and label input
	mechanisms,'' 2023.
	
	\bibitem{mirza2014conditional}
	M.~Mirza, ``Conditional generative adversarial nets,'' {\em arXiv preprint
		arXiv:1411.1784}, 2014.
	
	\bibitem{ronneberger2015u}
	O.~Ronneberger, P.~Fischer, and T.~Brox, ``U-net: Convolutional networks for
	biomedical image segmentation,'' in {\em Medical image computing and
		computer-assisted intervention--MICCAI 2015: 18th international conference,
		Munich, Germany, October 5-9, 2015, proceedings, part III 18}, pp.~234--241,
	Springer, 2015.
	
	\bibitem{chadha2024pytoacnn}
	K.~S. Chadha and P.~Kumar, ``{PyTOaCNN}: Topology optimization using an
	adaptive convolutional neural network in {Python},'' {\em arXiv preprint
		arXiv:2404.12244}, 2024.
	
	\bibitem{isola2017image}
	P.~Isola, J.-Y. Zhu, T.~Zhou, and A.~A. Efros, ``Image-to-image translation
	with conditional adversarial networks,'' in {\em Proceedings of the IEEE
		conference on computer vision and pattern recognition}, pp.~1125--1134, 2017.
	
	\bibitem{andreassen2011efficient}
	E.~Andreassen, A.~Clausen, M.~Schevenels, B.~S. Lazarov, and O.~Sigmund,
	``Efficient topology optimization in matlab using 88 lines of code,'' {\em
		Structural and Multidisciplinary Optimization}, vol.~43, pp.~1--16, 2011.
	
	\bibitem{kumar2023honeytop90}
	P.~Kumar, ``{HoneyTop90}: A 90-line {MATLAB} code for topology optimization
	using honeycomb tessellation,'' {\em Optimization and Engineering}, vol.~24,
	no.~2, pp.~1433--1460, 2023.
	
\end{thebibliography}
\end{document}